\documentclass[12pt]{article}
\usepackage{hyperref}
\usepackage{eepic}
\usepackage{graphicx}
\usepackage{amsmath,amssymb}
\usepackage[legacycolonsymbols]{mathtools}
\usepackage{here}
\usepackage{url}
\usepackage{ascmac}
\usepackage{bm}
\usepackage{caption}
\usepackage{multirow}

\title{Prediction of Toponium Levels Using\\ a Logarithmic Potential Model }

\author {Yasushi Muraki ${}^{\text {a)}}$ and Shoichi Shibata ${}^{\text {b)}}$\\ 
\normalsize a) Institute for Space-Earth Environment, Nagoya University,\\
\normalsize Chikusa, Nagoya, Aichi 464-8601, Japan\\ 
\small email: muraki@isee.nagoya-u.ac.jp\\
\normalsize b) Center for Muon Science and Technology, Chubu University,\\
\normalsize Kasugai, Aichi 487-8501, Japan\\
\small email:  shibata@isc.chubu.ac.jp\\}
\date{}

\begin{document}
\maketitle

\begin{abstract}
In this paper, the energy levels of the resonant states of toponium,
composed of top quark and anti-top quark,
are given on the basis of an empirical law.
We predict that the mass of the $n$-th resonant state of toponium is given
by ${\rm Mass}(n)=0.81 {\rm ln}(n)+347 \mathrm{\,GeV}$
from the empirical law on the resonance level of the bottomonium.
The cross-section produced by electron-positron collisions is
$3 \times 10^{-9}\mathrm{\,mb}$ and an electron-positron collider would need an energy of $270\mathrm{\,GeV} \times 270 \mathrm{GeV}$ to find out the resonance state of toponium. This prediction is based on the empirical law that the energy levels of hadron resonance states are expressed in logarithms. An interpretation of the appearance of quark resonance states in logarithmic intervals is also given in the paper. An application of this model, we present that the Okubo-Zwig-lizuka law can be viewed as a creation and annihilation problem of the two-dimensional resonance planes.
\end{abstract}

\section*{I. Introduction} -  Developments in Potential Models for Resonant States of Elementary Particles\\

Around 1960, many hadron resonance states were discovered through accelerator experiments. They are well represented by the Chew-Frautschi plot ${ }^{1-3)}$. The Chew-Frautschi plot is derived based on the Regge theory of hadron scattering ${ }^{4)}$. Recent studies on the Regge trajectory are detailed in the paper by Tang and Northbury (2000) ${ }^{5)}$.

Meanwhile, one of the authors was looking for the optimal spacing of the resonances that could reproduce the transverse momentum distribution of hadronic particles produced in high-energy hadron-hadron collisions. This process is now described as the "hadronization process" in which quark pairs form parton cascade showers, but at the time it was explained as the decay process of a meson "fire ball". The longitudinal momentum of the produced pions depends on the initial energy of the collision, but the transverse momentum of the secondary particles produced by the decay does not depend on the initial energy of the collision. This is because there is not necessary for a Lorentz transformation in the transverse direction.

In the case of a logarithmic distribution of resonance states, the transverse momentum distribution observed in this analysis was realized. We first hypothesized that the high-order resonance states in the Chew-Frautschi diagram were the initial "meson fireballs". This is similar to the process of photon cascade emission from the excited atom, but since pions have zero spin ( $S=0$ ) while photons have spin ( $S=1$ ), the selection rules for transitions from high-energy levels to low-energy levels are different. We investigated this transition process by solving the Schr$\ddot{o}$dinger equation for a quark-antiquark bound state under a logarithmic potential. We confirmed that the transverse momentum distribution of the produced hadron is well reproduced when the interval of the logarithmic scale is given ${ }^{6-7)}$. In modern words, we have obtained the transverse momentum of the pions produced in the parton cascade.

Therefore, we proposed that the quark-antiquark bound state is realized under the force of the inverse of the distance ( $\sim 1 / \mathrm{r}$ ) or the logarithmic potential $(\mathrm{V} \sim \ln (\mathrm{r}))^{6)}$. This logarithmic potential attracted an attention of theoretical physicists at the IHEP conference held in Tokyo around 1978 ${ }^{8-9)}$. This was after the discovery of the Upsilon resonance. The universality of the mass difference between the two heavy quarks, $\mathrm{M}\left(\Upsilon^{\prime}\right)$ - $\mathrm{M}(\mathrm{\Upsilon}) \approx \mathrm{M}\left(\psi^{\prime}\right)$ - $\mathrm{M}(\psi) \approx 0.56 \mathrm{GeV} / \mathrm{c}^{2}$, is naturally derived from the logarithmic potential.

However, the mass spectrum of charmonium (and bottomnium) cannot be reproduced by the Chew-Frautschi plot. Furthermore, the QCD potential predicts an energy interval of
$\rm{M}\left(\Upsilon^{\prime}\right)-\rm{M}(\Upsilon)
\approx(2/3)\left\{\rm{M}(\psi^{\prime})-\rm{M}(\psi)\right\}$,
which differs from the observed value by a factor of 2/3.
The details of this point are described in a 1977$'$s paper by Quigg and Rosner$^{10)}$ and further analysis has been carried out by them and presented in the reference ${ }^{11)}$. The review was made by Martin in 1991${ }^{12)}$ and following situation is well summarized in a paper by J-M Richard (2012) ${ }^{13)}$ which also includes a discussion of the $\ell-\mathrm{s}$ and s-s coupling effects in the charm and beauty spectroscopy by the logarithmic spacing.

In this paper, we first present the mass spectrum of the q\={q} state, which consists of light quarks, charm quarks, and bottom quarks. We then give an interpretation on the realization of the logarithmic-interval spectrum. Next, we discuss baryons consisting of three quarks. We also give an interpretation of the Okubo-Zwig-Iizuka rule ${ }^{14-16)}$ from a different perspective. Finally, we give the mass spectrum of the toponium ${ }^{17)}$, which is composed of the top quark and anti-top quark. We show that an electron-positron collider with an energy of about 250 GeV is necessary to test whether this prediction is correct or not.

\section*{2. Mass Plot of Particle Resonances}

\section*{2-1. Mass plot by the Chew-Frautschi plot}

In this section, we first consider whether the resonance states of the J/$\Psi$ and $\Upsilon$ particles, which are composed of charm and bottom quarks, can be expressed using the Chew-Frautschi plot.
First, the orbit of the $\rho$ meson, which is a resonance state of light quarks, is shown in Figure 1a (we will call this orbit the "trajectory of family I"). Figure 1a (top) is a plot of total angular momentum ($\it J$) on the vertical axis and the squared mass ($\it s$) on the horizontal axis. The four resonance states used in the plot are the four mesons $\rho$, A2, $\rho 3$ and a4, and you can see that the four resonance points are well plotted on a very clear straight line. This means that the straight line proposed by Chew and Frautschi reproduces the experimental data points very accurately. The correlation coefficient $R$ between the straight line and the data points is 0.9997. The correlation is extremely good.

On the other hand, the charmonium (family II) and bottomonium (family III) are formed from heavy quarks and their antiparticles. The plots are shown in Figures 1b (middle) and 1c (bottom). The experimental data points for families II and III in Figures 1b and 1c cannot be represented by the Chew-Frautschi plot. The data points suggest that they are represented by an exponential function. The numerical values of each data point are shown in the third column of Table I. The values in the third column of the first family correspond to the slope and intercept of the Chew-Frautschi plot.

\section*{2-2. Mass plot on the logarithmic scale for mesons}

As we have seen above, the Chew-Frautschi plot is a very good approximation for describing the resonance states composed of light quarks in the first family, but it is not suitable for describing the charmonium and bottomonium, which are composed of heavy quark pairs. Is there a formula that can describe these three families of bosons in a unified way? We investigated these resonance states using the logarithmic spacing model proposed by one of the authors in $1968^{6)}$. The results are shown in Figure 2.

Figures 2a, 2b, and 2c are plots corresponding to the resonance states of families I, II, and III, respectively. The values of the slopes, intercepts, errors, and correlation coefficients are summarized in Table 2. The correlation coefficient $R$ is an indicator of the degree to which the data points are reproduced by a straight line when the data are plotted as a straight line on a semi-logarithmic plane. In the case of the upsilon family, the fitting curve is close to 1.0 with 4 -digit precision ( 0.9999 ), indicating that it fits the data almost perfectly. These are shown in the fourth column of Table 2. For $\Upsilon_{5}$ (10753), $\Upsilon_{6}$ (10885), and $\Upsilon_{7}$ (11020), our empirical curve predicts correct resonance masses as $\Upsilon_{5}(10754 \pm 4), \Upsilon_{6}(10900 \pm 4)$, and $\Upsilon_{7}(11023 \pm 4 \mathrm{MeV})$ respectively.

\section*{2-3. Mass plot on the logarithmic scale for baryons}

In the previous section, we investigated the resonance states of mesons formed by quark-antiquark systems. The next question is whether the logarithmic spacing model can be applied to the three-body systems of baryons. Therefore, we investigated the resonance states of the $\mathrm{N}^{*}$ and $\Delta$ baryon systems formed by light d-quarks and u-quarks. We report here the results of this study. The baryon family consists of three quarks, so their dynamical motion is not as simple as that of mesons. For example, do qq and q move as a pair like mesons, or do the three quarks rotate with gluons connecting them at $120^{\circ}$ intervals at equal intervals? There are various possibilities. For details, please refer to the paper by Jean-Marc Richard (2012)$^{13)}$.

Figure 3 shows the observed $\mathrm{N}^{*}$ and $\Delta$ baryon resonance states plotted on a semi-logarithmic plane classified by mass, total angular momentum ($\it J$), and parity ($\it P$). In the Chew-Frautschi plot, angular momentum is plotted every other point, but in our model, they are plotted in order of quantum number. This is the main difference between the Chew-Frautschi plot and ours. Detailed analyses of the Chew-Frautschi plot have been carried out for baryons by, for example, Tan and Norbury (2000)${ }^{18)}$ and Klempt and Metsch (2012)${ }^{19)}$. For a detailed discussion, please refer to these papers. Here, we discuss only characteristic baryon resonance states.

The baryon resonances used in this paper are the five resonance states
in the $\rm{N}^{*}(\rm{~J}=1/2^{+}$ ) series
and the three (parity +) and four (parity -) baryons
in the $\Delta$(3/2) family.
The masses of the specific resonance states used in the analysis are
(938, 1440, 1710, 1880, 2100 MeV)
for the former baryons, and their total angular momenta ( $J$ )
of the resonance states are all 1/2, and the parities are all +.

On the other hand, the latter baryons are
$\Delta$(1,232(3/2$^{+}$), 1,950(7/2$^{+}$), 2,420(11/2$^{+}$))
and four delta baryons
$\Delta$(1,700(3/2$^{-}$), 1,930 (5/2$^{-}$), 2,200 (7/2$^{-}$), 2,400 (9/2$^{-}$)).
These are plotted
in Figures 3a for 1/2$^{+}$ 
family and 3b for $\Delta$ family, respectively.
When each family is plotted on a semilogarithmic plane,
the correlation factor $R$ shows how well they are represented by a straight line and the value of $R$ is shown in the fourth column. The correlation coefficient $R$ is described by the range of 0.994 to 0.997 , which is close to the value of $R$ for the $\rho$ meson family, 0.9987.

\section*{3. Discussion and Interpretation}

\subsection*{3.1 Plotting and Interpretation}
What does the logarithmic spacing of the resonance states mean?
In this section, we will discuss what can be derived from these plots.\\

(i) First, the fact that the energy levels of the resonance states are realized in the logarithmic intervals suggests that the force between quarks and antiquarks may be 
expressed as inversely proportional to the distance ( $\mathrm{F} \propto 1 / \mathrm{r}$ ). In other words, since the derivative of the force is related to the potential energy $(\mathrm{V} \propto \ln (\mathrm{r})$ ), it is expected that the 
energy levels of the resonance states will also be preesented by the logarithmic functions, since the kinetic energy is expected to be narly constant. An example of such a potential is the Coulomb field in a two-dimensional plane, which is expressed as a Gaussian integral for the electric field in a plane. 
Under a constant $\it r$ , 
$\int$$\int$ $E_{r}$ d $\phi$ dr = 2 $\pi$ r $\cdot$ $E_{r}$ = q. 
This leads to $E_{r}$=q / 2 $\pi$r 
and V=q / 2 $\pi$$\cdot$$\it ln (r)$.\\

(ii) If the force between quarks and antiquarks can be approximated by $\mathrm{F} \propto 1 / \mathrm{r}$, it can be seen that only circular motion is allowed in the force field. In addition, it can be seen from the Virial theorem that the kinetic energy of the quark is constant, $<\mathrm{mv}^{2}>\approx$ constant. Furthermore, since the angular momentum $L$ is limited by $L=\mathrm{p} \times \mathrm{r}=\mathrm{p} \times \mathrm{r}=\mathrm{mvr}$, this leads to an important relation that the radius $<r>$ of each resonance state is proportional to the angular momentum $L$. We confirmed the above assertion by numerically solving the Schrödinger equation under a logarithmic potential ${ }^{7)}$. This state may be depicted in Fuigure 4.

Here we regard that $\ell$ (or $L$ ) and $\it n$ are wave numbers that undulate in a narrow region. Figure 4 illustrates the wave situation. It shows that the circle where the wave is excited increases with the wave number proportionally. The unit length ( $\ell_{0}$ ) is considered to be 0.25 Fermi. Therefore, we think that baryon resonance states with different angular momenta ( $\ell$ ) and principal quantum numbers ($\it n$) may have the common horizontal axis and can be plotted on the same graph. The slopes of these lines are almost identical within the range of experimental error, which is consistent with this assumption. For this reason, in Figure 3, the physical quantity on the horizontal axis is expressed as ( $\mathrm{n}+\ell$ ).\\

(iii) From above discussions, we assume here when a quark and an antiquark are in motion with a pair, their motion can be described as motion in a two-dimensional plane. As a result, we discuss that the principal quantum number ($\it n$) and angular momentum $(\ell)$ of the resonance state can be treated equivalently. The equation of motion of a particle in a centrifugal force field in the direction $r$ is
 described in both classical and quantum mechanics as the motion of a particle under an effective potential. If the effective potential is 
 now $V_{eff}$,  the Schr$\ddot{o}$dinger equation in the three dimensional space can be written as\\

$\hbar^{2} / 2 \mu \cdot \ddot{u}(\mathrm{r})+[\mathrm{E}-\mathrm{Veff}] u(\mathrm{r})=0$\\
      
       and\\
       
$V_{eff }$ = V(r) + \{$\hbar^{2}$ $\ell$($\ell$+1)\} / 2${r}^{2}$\\

where $u(\mathrm{r})$ is the wave function in the two dimensional space.\\

(iv) However, the motion of a particle in a two-dimensional plane possesses only a $J z$ component as angular momentum, and the eigenvalue $J z$ takes simple integer values of $J z= \pm 1,2,3, \ldots$. The angular momentum ( $\ell$ ) does not increase $\ell^{2}+\ell$ with $\ell$ as in the centrifugal force potential under 3D motion. The motion of the particle under our proposed logarithmic potential is expected as to be a circular motion and the mean value of its orbital radius is uniquely determined by the quantum number. For the principal quantum number $n$ and angular momentum $\ell$, the orbital radius varies slightly with the mass of the constituent quarks, but the orbit is circular for both $n$ and $\ell$,
so the logarithmic argument is taken as $(n+\ell)$. 
In other words, $n$ and $\ell$ are quantum numbers of the same rank for the motion of the quark pair and they represent the masses of resonance levels.  The wave function predicted by the Shr$\ddot{o}$dinger equation in the two dimensional space$^{20)}$ is given in Appendix 1.

In the paper, Quigg and Rosner predicted that the resonance level is expressed as $\mathrm{E}_{\mathrm{n}}=\mathrm{C} \cdot \ln [\mathrm{A} \cdot(n+\ell / 2-1 / 4)]$. This leads to, $\left(\mathrm{E}_{3}-\mathrm{E}_{2}\right) /\left(\mathrm{E}_{2}-\mathrm{E}_{1}\right) \approx$ $(\ln 11 / 7) /(\ln (7 / 3))=0.533$ for resonance levels 1, 2, and 3$^{12)}$.
In our model, on the other hand, this value gives $(\ln 3-\ln 2) /(\ln 2-\ln 1)=0.586$. The Upsilon resonance state has three states, $\Upsilon 1, \Upsilon 2$ and $\Upsilon 3$, determined with very good accuracy. The above ratio for the experimental data is 0.589 . The agreement between our predictions and the experimental data is very good.\\

(v) In 1962, in a paper, Chew and Frautschi made an important point. The following is a quotation from their paper: "The only constants added to $h$ and $c$ are constants with a length (or mass) dimension, not arbitrary parameters. Elementary particles do not exist." The authors of this article do not agree with the last statement that elementary particles do not exist, but based on current experimental results, it is true that when quarks approach each other in the elementary region (actually $1 / 4$ fermi or $2.5 \times$ $10^{-14} \mathrm{~cm}$) of 25 femto meter, a resonance states are formed. The fact that the energy states of not only the $\mathrm{u}, \mathrm{d}$, and s quarks, but also the c and b quarks, can be described by curves with almost the same slope suggests the existence of interactions corresponding to such "fundamental lengths". This characteristic length $\Delta \ell$ is derived from the uncertainty relation $\Delta \ell \approx(\mathrm{h} / 2 \pi) \cdot \mathrm{c} / \Delta \mathrm{E}$ and $\Delta \mathrm{E}=810 \mathrm{MeV}$.

If there were a space with a dimension of fundamental length, the oscillation of the resonance state of elementary particles could be described as in Figure 5. In this case, $r$ would be constant and $v$ would be quantized. Then from the Virial theorem $v$ would not be constant and it would not be described by a logarithmic potential model. Note that the experimental data are limited to a finite number of resonance states, although it is thought that many quantum states are allowed by constant space $(r)$ model. Because when the space region is narrow, many higher resonances would be formed in the potential well.\\

(vi) The slopes of $\Delta$ and $\mathrm{N}^{*}$ baryons are less steep than that of the meson. Since the error for $\Delta$ is large, if we focus on the $\mathrm{N}^{*}$ and $\Upsilon$ particles, we observe a discrepancy of about $3 \sigma$ between the two slopes. If the quarks in the baryon move in pairs of qq-q, the ratio predicted from the reduced mass should be $4 / 3=1.33$. However, as shown in Figure 6, the ratio is 0.88 , and it seems unlikely that the resonance state is realized by a qq-q pair.

\subsection*{3.2 An Interpretation of the Okubo-Zwig-Iizuka rule}

Present hypothesis that a resonance state of elementary particles is realized when the distance corresponding to the energy of a quark-antiquark pair is realized near a characteristic wavelength, and that the motion of the particle can be expressed in terms of two-dimensional plane, may provide a dynamic explanation of the OZI rule. The following is a discussion of this point.

The left side of Figure 7 shows the decay process of the c\={c} resonance state $\psi(3770)$, which is depicted by quark lines. As is known from the OZI rule, this process decays through strong interactions, because the quark line does not disappear. Therefore, the $\it R$ ratio for $\Psi(3770)$ in the e$^{+}$e$^{-}$ type accelerators will be wider than the peaks 
for $\mathrm{J} / \psi(3096)$ and $\psi(2\mathrm{S})$. 
On the other hand, in the decay process on the right side of Figure 7, the charm
quark line must disappear once to create a new quark pair, so the production of J/$\psi$ particles is suppressed. This is an explanation on the Okubo-Zwig-Iizuka rule$^{14-16)}$.

If we think of the production and annihilation of particles as a motion on a two-dimensional plane, the decay process on the left side is allowed because it is a decay process on the same plane, and the decay width of $\psi(3770)$ widens to 27 MeV. On the other hand, in the decay process of the $\mathrm{J} / \psi$ particle on the right side, it is necessary for the charm quark line to disappear for a moment. Therefore, the initial plane where the charm quark pair existed must be erased, and a new plane corresponding to the new quark pair must be prepared. The OZI rule can be interpreted as a problem of continuity/discontinuity of motion in this two-dimensional plane. This is probably why the J/$\psi$ particle has a long lifetime and a very narrow decay width such as 93 keV.

\subsection*{3.3 The trajectories of $\rho$ mesons and $N^{*}\left(1 / 2^{+}\right)$resonances}

In this section, we will first explain Tables 2 and 3 in detail. The value {\it R} shown in the fourth column of Tables 2 and 3 is a correlation coefficient. In other words, it is an indicator of how well the line represents the mass values when the mass values of the resonance states are plotted on a semilogarithmic plane by a straight line. The values in the second and third columns show the slope of the line on the logarithmic scale and the intercept on the vertical axis. 
This intercept corresponds to $\it Ln$(1)=0, which is very close to the first
 levels of the J/psi and Upsilon resonances of charm and bottom particles, 3096.9 MeV and 9460.3 MeV respectively. 
 The correlation coefficient {$\it $} between the data and the straight line for the bottomonium line is 0.99997, which is a value that is close to 1.0. 
 This suggests that the resonance order of elementary particles can be described by a logarithmic spacing model.

The numbers in parentheses in the second and third lines are the slope and intercept of the line obtained by the least-squares method, taking into account the error on the energy of each resonance state. The values of each slope and intercept are shown with their respective errors. The errors are smaller for families II and III, which are formed from heavy quarks.

Furthermore, the correlation coefficient $R$ for family III is closer to 1.0 than the correlation coefficient for family II, by about one order of magnitude. The error in the correlation coefficient $R$ is calculated assuming that the deviation from the straight line follows a Gaussian distribution, and the value 0.99997 is evidence that this assumption is correct.

However, for the resonance states and baryons in Family I, which are composed of light quarks, the error varies greatly depending on how the error is evaluated and added. For example, in the case of the orbitals of Family I, which consists of four resonance states ($\rho, a 2, \rho 3$, a4), the slope and intercept differ greatly depending on whether ($\it a$) half the full width of the resonance $(\Gamma / 2)$ is adopted or ($\it b$) the error on the mass measurement is adopted. The slopes and intercepts for cases ($\it a$) and ($\it b$) are (839.75 $\pm$ 88.23) and (753.78 $\pm 67.88$ ) and (798.2 $\pm 1.11$ ) and (774.93 $\pm 0.25$ ), respectively. Therefore, the range of errors in the resonance trajectory composed of light quarks differs greatly depending on the method of processing measurement errors and the use of the least-squares method, but the variation in the median is not so large.

The same is true for baryons, with the $\mathrm{N}^{*}\left(1 / 2^{+}\right)$baryon series including the proton at the lowest rank with a mass of 938.3 MeV . This mass has been accurately measured to six decimal places. Klempt and Metsch$^{19)}$ solved this problem by assuming a measurement error of $938 \pm 30 \mathrm{MeV}$ for the proton and derived the slope. We assumed two cases for the measurement error for the 938 MeV proton: $(\alpha) \pm 20 \mathrm{MeV}$ and $(\beta) \pm 1 \mathrm{MeV}$, and used the least-squares method to calculate the slope and intercept of the line for each case. As a result, the values $(\alpha)(735.6 \pm 47.0) \mathrm{MeV},(914.8 \pm$ 55.3) MeV , and $(\beta)(711 \pm 21) \mathrm{MeV},(936 \pm 19) \mathrm{MeV}$ were obtained for the five data points, respectively.

In Figure 8, we present the binding state of the rho meson family (green), charmonium state (red) and botomnium (blue) taking into account OZI decay effect. The light blue region corresponds to quasi stable region of these resonances below the flavor threshold.

\section*{4. Production of Toponium by Electron-Positron Collider}

The top quark ($\it t$) was discovered in hadron colliders, and it is much heavier than $\mathrm{u}, \mathrm{d}, \mathrm{s}, \mathrm{c}$, and b quarks, with a mass of $173 \mathrm{GeV} / \mathrm{c}^{2} .^{21-24)}$ It is an interesting question whether such a top quark forms a resonance state with its antiparticle. However, this question has not yet been proven experimentally. Let's consider here that such heavy quarks can be bound by gluons and form bound states, toponium ($\it T$). In addition, their motion in a narrow space follows a logarithmic potential.

Then we can predict the resonance levels of toponium. To prove this hypothesis, an electron-positron collider with an energy of $270 \mathrm{GeV} \times 270$ GeV or more is needed. Currently, three projects (ILC, BEPC, CLIC) have been proposed, and we hope that they will be realized as soon as possible. Actually the CLIC proposal ${ }^{25)}$ describes that at first stage they will construct an electron and positron collider with the energy of 190 GeV $\times 190 \mathrm{GeV}$. Therefore, the possibility of the discovery of toponium, if they will really stay in the resonance states.

The mass of the top quark is known to be very heavy from hadron machine experiments ${ }^{26)}$. Therefore, the lowest mass of the resonance state formed in pairs with antiparticles is expected to be about 347 GeV . If such heavy elementary particles are bound in the central region of the logarithmic potential, many resonance states are expected to appear up to the maximum mass of $524.5 \mathrm{GeV}=177.5 \mathrm{GeV}+347 \mathrm{GeV}$, which can decay into
( $W^{+}bW^{-}\bar{b}$ ) according to the OZI rule. However, for higherorder resonances, the energy order between the resonance states becomes narrower, and it is expected that a continuous aspect will appear within the range of measurement error. If the mass is estimated according to the current logarithmic interval model, the mass of the higher-order resonance of toponium is around $T(80)$, and the mass is expected to be about 350.54 $\mathrm{GeV} / \mathrm{c}^{2}$. Here, the mass of $\it T$(1) is assumed to be $347.0 \mathrm{GeV} / \mathrm{c}^{2}$. In the calculation, the same slope of 0.81 GeV as for the Upsilon family is applied to the slope of the logarithmic interval for toponium.  However above assumption holds in a case that top quark will not make mesons including light quarks like \={u} or \={c}.  In the latter case, we will see only a few sharp resonance states near 347 GeV.

Above $\mathrm{n} \geq 80$, the interval between the mass $(\mathrm{n}+1$ ) and the mass ( n ) becomes narrower than 10 MeV , and it will probably be very difficult to identify each of them experimentally. Experimental results would show that the spectrum increases rather continuously up to 524.5 GeV . It is sure that next $\mathrm{e}^{+} \mathrm{e}^{-}$colliders will work as Top factory.

However, in reality, excited states with such high quantum numbers of $\mathrm{n}=80$ would not be observed, because the orbital radius would expand to $2 \times 10^{-13} \mathrm{~cm}$ at $\mathrm{n}=80$. It is not sure whether a heavy top quark can be bound by a gluon to such a long distance. It is near the radius of Helium nucleus.

From the above discussions, the energy of the electron-positron collider must be able to accelerate electrons and positrons each other to 270 GeV . The production cross section of the lowest energy toponium is expected to be $7 \times 10^{-9} \mathrm{mb}$ (baseline). In Figure 9
, we present the expected cross-section to the production of toponium for future electron and positron collider.

\section*{5. Summary}

In this paper,\\

(1) We presented that the Chew-Frautschi plot does not reproduce the resonance states of charmonium and bottomonium, but that the logarithmic spacing plot does.\\

(2) We then applied the logarithmic spacing model to the resonance states of toponium, which are expected to be confirmed in future electron positron collider experiments, and predicted what the mass spectrum would be like.\\

(3) The fact that hadron resonance states can be expressed using the logarithmic spacing model strongly suggests that the c\={c} and b\={b} resonance states are excited at integer multiples of the unit wavelength in a small region of space.\\

(4) Assuming that the quark-antiquark resonance state is excited in a two dimensional plane, it is suggested that the OZI rule can be interpreted from a geometric perspective.
\\
\section*{Acknowledgement}

The authors thank to Chubu university and Nagoya university for providing us their computer and scientific facilities. \\

\section*{Table }

  Table 1\\
  \\
\begin{tabular}{|l|l|l|l|}
\hline * &  Fitting parameter & Mass square(s) -quantum number & Family  \\
\hline Rho trajectory & $\mathrm{J}=0.9128 \mathrm{s}-0.437$ & s=1.096$\cdot$J+0.479 [GeV] &I \\
\hline $J/\psi$ trajectory & n=0.2145$\exp(0.1623s)$ &   

\begin{tabular}{r} 
$\ln (n)=0.1623 \cdot s -1.539 $ \quad  or \\
$s=6.161\cdot ln(n) $+ 9.482  
\end{tabular} & II \\

\hline $\Upsilon$  trajectory & $n=0.0041 \exp (0.0615s)$ & 

\begin{tabular}{r}
$\ln (n)=0.0615 \cdot s - 5.497 \quad$ or \\
$s=16.26\cdot ln(n)$ + 89.38
\end{tabular} & III \\
\hline
\end{tabular}
\\

Table 2\\
\begin{center}
\begin{tabular}{|l|l|l|l|l|}
\hline
Series & Slope & \begin{tabular}{l}
Intercept \\
(MeV) \\
\end{tabular} & R (data point) & Family \\
\hline
(I) Rho trajectory & \begin{tabular}{l}
$860 \pm 30$ \\
$(798.2 \pm 1,1)$ \\
\end{tabular} & \begin{tabular}{l}
$752 \pm 29$ \\
$(774.9 \pm 0.3)$ \\
\end{tabular} & \begin{tabular}{c}
0.9987 (4) \\
(4 points) \\
\end{tabular} & I \\
\hline
(II) J/ $/$ trajectory & \begin{tabular}{l}
$828 \pm 24$ \\
$(849.0 \pm 0.3)$ \\
\end{tabular} & \begin{tabular}{l}
$3105 \pm 23$ \\
$(3096.9 \pm 0.1)$ \\
\end{tabular} & \begin{tabular}{c}
0.9991 (4) \\
(6 points) \\
\end{tabular} & II \\
\hline
(III) $\Upsilon$ trajectory & \begin{tabular}{l}
$809.2 \pm 4.2$ \\
$(810.50 \pm 0.39)$ \\
\end{tabular} & \begin{tabular}{l}
$9460.6 \pm 4.0$ \\
$(9460.90 \pm 0.25)$ \\
\end{tabular} & \begin{tabular}{c}
$0.99997(4)$ \\
(7 points) \\
\end{tabular} & III \\
\hline
\end{tabular}
\end{center}

Table 3

\begin{center}
\begin{tabular}{|l|l|l|l|l|}
\hline
Series & Slope & \begin{tabular}{c}
Intercept \\
$(\mathrm{MeV})$ \\
\end{tabular} & R (data point) & J and parity \\
\hline
(I) N* resonances & \begin{tabular}{c}
$735.56 \pm 38.2$ \\
$( \pm 47)$ \\
\end{tabular} & \begin{tabular}{c}
$921.4 \pm 38.2$ \\
$( \pm 55)$ \\
\end{tabular} & \begin{tabular}{c}
$0.9965(6)$ \\
$(5$ points) \\
\end{tabular} & All $1 / 2+$ \\
\hline
\begin{tabular}{c}
(II) $\Delta$ resonances \\
(parity + ) \\
\end{tabular} & \begin{tabular}{c}
$721.78 \pm 77.5$ \\
$( \pm 23)$ \\
\end{tabular} & \begin{tabular}{c}
$1209.1 \pm 87.2$ \\
$( \pm 20)$ \\
\end{tabular} & \begin{tabular}{c}
$0.9942(3)$ \\
$(3$ points) \\
\end{tabular} & $3 / 2+, 7 / 2+, 11 / 2+$ \\
\hline
\begin{tabular}{c}
(III) $\Delta$ resonances \\
(parity -) \\
\end{tabular} & \begin{tabular}{c}
$753.81 \pm 55.7$ \\
$( \pm 56)$ \\
\end{tabular} & \begin{tabular}{c}
$1162.8 \pm 69.5$ \\
$( \pm 42)$ \\
\end{tabular} & \begin{tabular}{c}
$0.9945(4)$ \\
$(4$ points) \\
\end{tabular} & \begin{tabular}{l}
$3 / 2-, 5 / 2-, 7 / 2-$, \\
$9 / 2-$ \\
\end{tabular} \\
\hline
\end{tabular}
\end{center}

\section*{References}
1)  G. F. Chew and S.C. Frautschi, Physical Review Letters, 5 (1960) 580.
Unified approach to high-and low-energy strong interactions on the basis of the Mandelstam representation.

 DOI:\href{https://doi.org/10.1103/PhysRevLett.5.580}{https://doi.org/10.1103/PhysRevLett.5.580}\\
G. F. Chew and S.C. Frautschi, Physical Review, 123 (1961) 1478. Dynamical theory for strong interactions at low momentum transfers but arbitrary energies,

DOI: \href{https://doi.org/10.1103/PhysRev.123.1478}{https://doi.org/10.1103/PhysRev.123.1478}\\
2) G. F. Chew and S.C. Frautschi, Physical Review Letters 7 (1961) 394. Principle of equivalence for all strongly interacting particles within the S-matrix framework

DOI: \href{https://doi.org/10.1103/PhysRevLett.7.394}{https://doi.org/10.1103/PhysRevLett.7.394}\\
3) G. F. Chew and S.C. Frautschi, Physical Review Letters, 8 (1961) 41, Regge Trajectories and the principle of maximum strength for strong interactions. 

DOI: \href{https://doi.org/10.1103/PhysRevLett.8.41}{https://doi.org/10.1103/PhysRevLett.8.41}.\\
4) T. Regge, Nuovo Cimento 14 (1959) 951, 18 (1960) 947. 

\href{https://doi.org/10.1007/BF02728177}{https://doi.org/10.1007/BF02728177}\\
5) A. Tang and J. W. Norbury, Physical Review D62 (2000), 016006.

 \href{https://doi.org/10.1103/PhysRevD.62.016006}{https://doi.org/10.1103/PhysRevD.62.016006}\\
6) Y. Muraki, Prog. Theo. Phys. 41 (1969) 473. A new mass formula of elementary particles, 

DOI: 10.1143/PTP.41.473. \\ ${  }$Y. Muraki, Soryusiron Kenkyuu Electronnics Vol.38-1 (1968) 34-42 in English. 

 https://doi.org/10.24532/soken.38.1.34\\
7) S. Iida and Y. Muraki, Proceed. 11 ${ }^ {th}$ of Int. Conf. on Cosmic Rays (Budapest) (1969).

Acta Physica Hungarica Supplement 29 pp 271-277 (1970).\\
8) J.D. Jackson, C. Quigg, and L.L. Rosner, New Particles, Theoretical

 Proceeding of the $19^{th }$ International Conference on High Energy Physics (Tokyo) 1978 B8. (ICHEP78), CERN Document Server p391, New Particles, Theoretical, by Jakson, David ..\\
9) M. Machacek and Y. Tomozawa, Psi Phenomenology and nature of Quark Confinment, Annals of Physics, 110 (1978) 407-420.

http://dx.dvi.org/10.1016/0003-4916(78)90037-4\\
10) C. Quigg and J. L. Rosner, Quarkonium Level Spacings, Physics Letters B71 (1977) 153. 

DOI: 10.1016/0370-2693(77)90765-1\\
11) C. Quigg and J. L. Rosner, Quantum Mechanics with Applications to Quarkonium, Physics Reports, 56 No. 4 (1979) 167-235.

https://doi.org/10.1016/0370-1573(79)90095-4\\
12) A. Martin, Potential Models of Hadrons, Nuclear Physics B (Proc. Suppl.) 23B (1991) 316-327.

https://doi.org/10.1016/0920-5632(91)90698-E\\
13) Jean-Marc Richard, An introduction to the quark model, arXiv:1205.4326v2, [hep-ph] 24 May 2012.\\
14) S. Okubo, Physics Leters 5(2) (1963) 165-168. 

doi:10.1016/s0375-9601(63)925-48-9\\
15) G. Zwing (1964) CERN report No.8419/TH412(Report)\\
16) J. Iizuka, Progr.Theor.Phys. Suppl. 37-38(1966) 21-34. 

\href{https://doi.org/10.1143/PTPS.37.21}{https://doi.org/10.1143/PTPS.37.21}\\
17) E. Eichten and K. Gottfried, Physics Letters B66 (1977) 286.

\href{https://doi.org/10.1016/0370-2693(77)90882-6}{https://doi.org/10.1016/0370-2693(77)90882-6}\\
18)A. Tang and J.W. Norbury, Physical Review D62 (20009 016006.

https://doi/10.1103/PhysRev.D.62.016006\\
19) E. Klempt and B. Ch. Metsch, Multiplet classification of light-uark baryons,
Eur. Phys. J. A 48(2012) 127

DOI:10.1140/epja/i2012-12127-1\\
20) O. Atsbek, C. Deutsch, and M. Lavaud, 
Physical Review A9 (1974) 2617. Schr$\ddot{o}$dinger equation for the two-dimensional Coulomb potential\\
21) F. Abe et al. (CDF collaboration), Phys. Rev. Lett., 74 (1996) 2626.
 
https://doi.org/10.1103/PhysRevLet.742626\\
22) S. Abachi et al. (D0 collaboration), Phys. Rev. Lett. 74 (1996) 2632-2637.

 DOI:10.1103/PhysRevLett. 742632\\
23) M. Aabound et al. (ATLAS collaboration) Eur. Phys. J. C79 (2019) 290. 

arXiv:1810.01772 [hep ex]\\
24) V.K. Hachatryan et al., Phys.Rev. D93 7 (2016) 072004. 

arXiv:1509.04044.\\
25) O. Brunner et al., The CLIC project, 

ar Xiv:2203.09186v2 [physics.acc-ph] 18 Apr 2022\\
26) Particle Data Group, Progr. Theor. Exp. Phys. 2020 083c01 (2020) 741-761. 

DOI:10.1093/ptep/ptaa104


\begin{figure}[t]
\centering
\includegraphics[keepaspectratio, width=10cm]{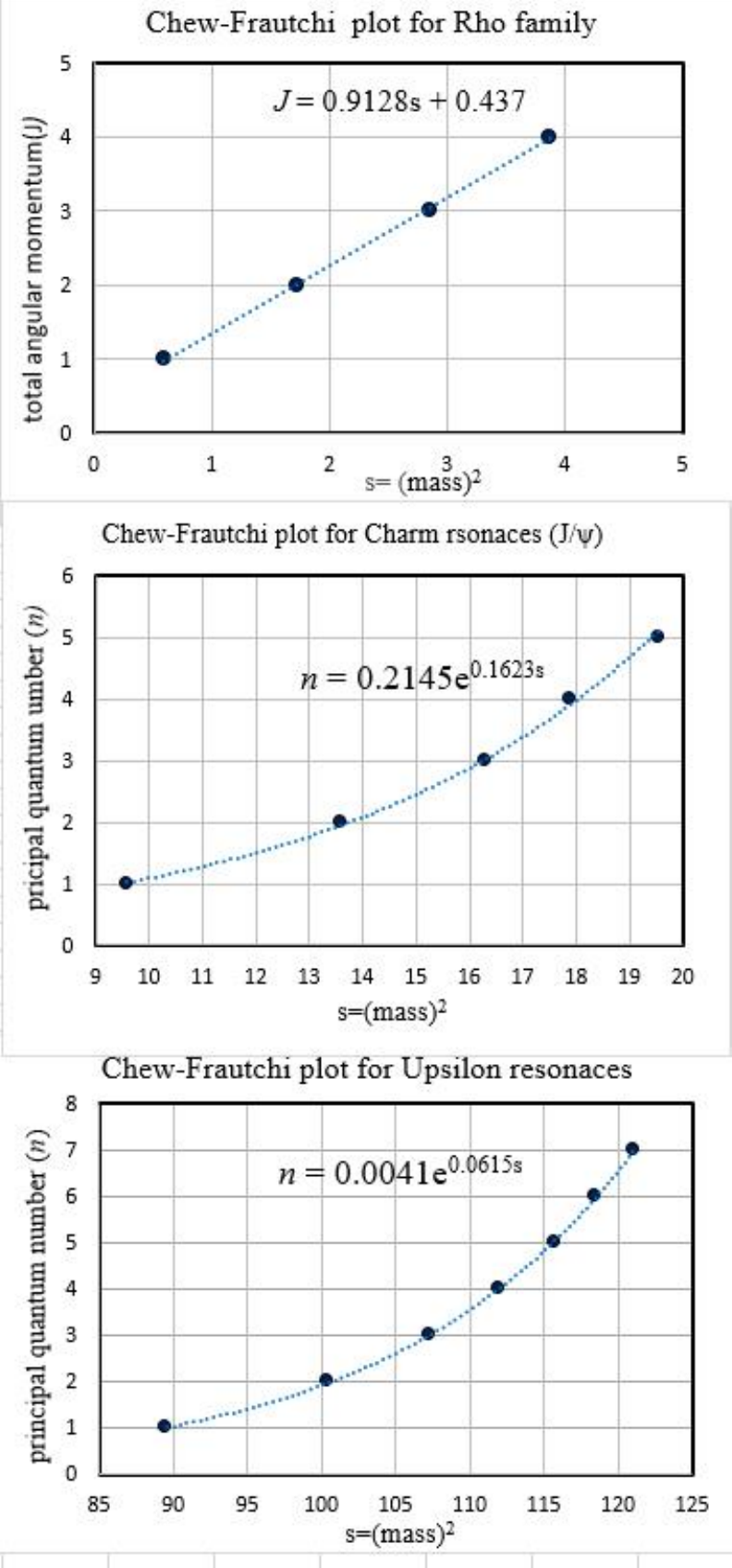}
\caption{The rho-meson trajectory, from top to bottom,  (a) charmonium (b) and botomonium state (c) are presented on the Chew-Frautschi plot. The horizontal axis presents the square of resonance mass, while the vertical axis corresponds to the angular momentum ($\it J$) or principal quantum number ($\it n$). For the charmonium and the botomonium states, these resonances are not well fit on the Chew-Frautschi plot, but rather well depicted by the exponential term.}
\label{fig:01}
\end{figure}

\begin{figure}[t]
\centering
\includegraphics[keepaspectratio, width=9.5cm]{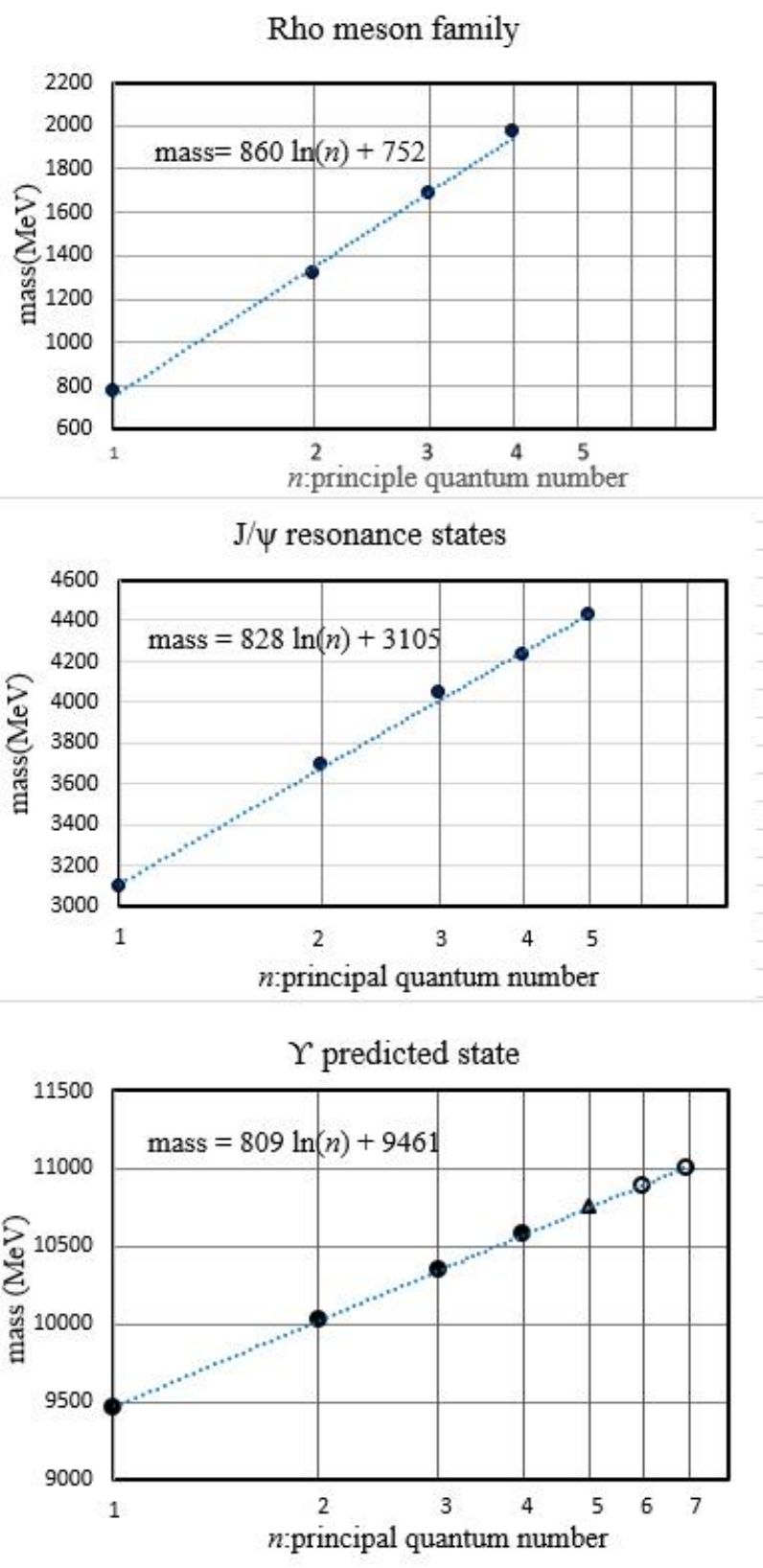}
\caption{The three meson resonance families are plotted on the semi-log diagram. All three trajectories fairly well plotted on the straight line. The horizontal axis corresponds to either angular momentum ($\ell$) or the principal quantum number ($\it n$). Vertical axis corresponds to the mass of each resonance. The slope of the straight line and the intercept is shown in each panel.}
\label{fig:02}
\end{figure}

\begin{figure}[t]
\centering
\includegraphics[keepaspectratio, scale=0.6]{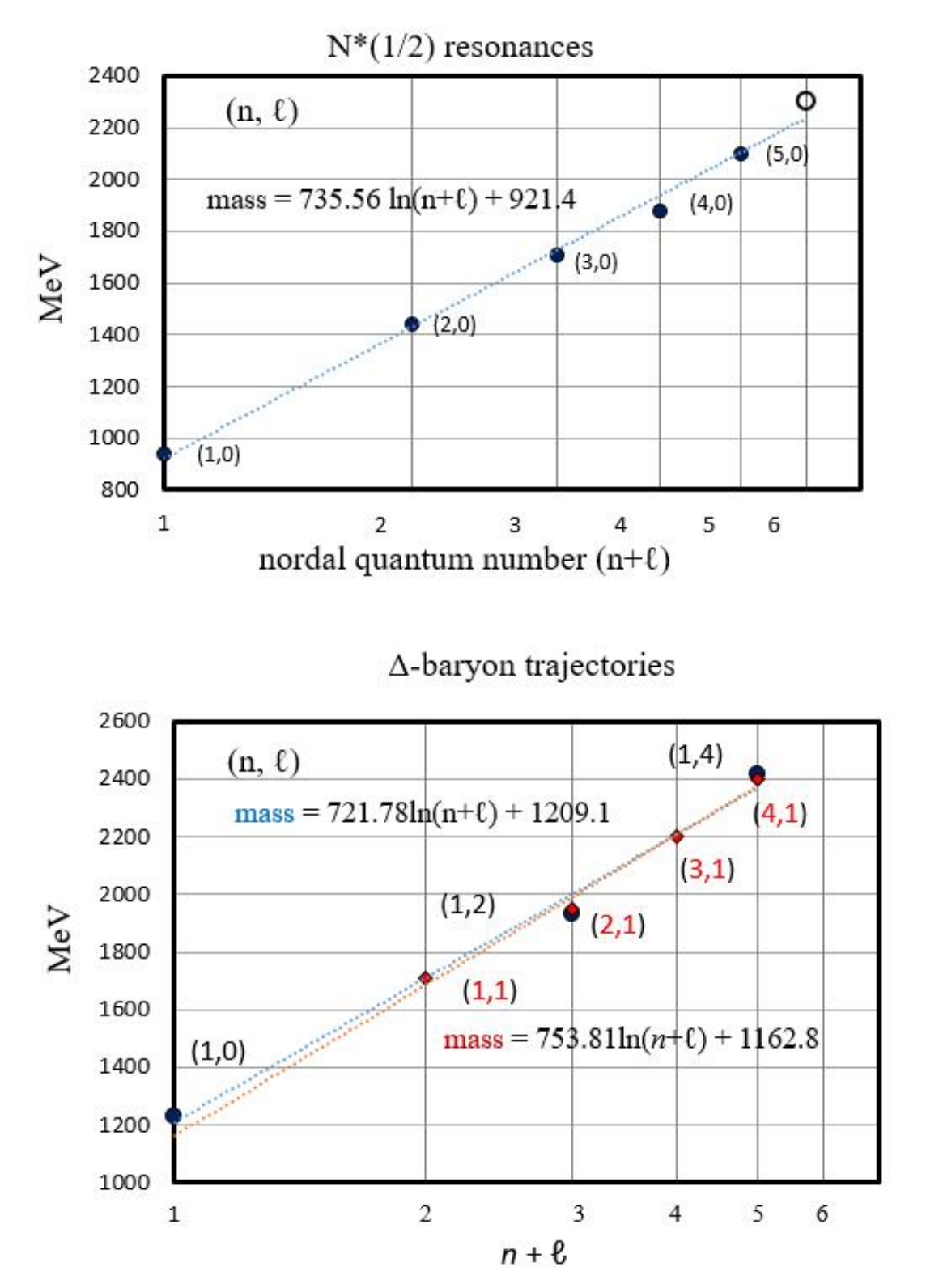}
\caption{Top figure shows the logarithmic plot
for the nucleon resonances of $N^{*}$(1/2) family.
The number of the bracket shows the principle quantum number ($n$) and the angular momentum ($\ell$). This case ($\it N^{*}$1/2), all the angular momentum are zero.  The bottom figure shows the logarithmic plot for the $\Delta$ baryon series.  The slope of each curve is given in the figure.}
\label{fig:03}
\end{figure}

\begin{figure}[h]
\centering
\begin{minipage}[t]{0.49\columnwidth}
    \centering
    \includegraphics[width=1.1\columnwidth]{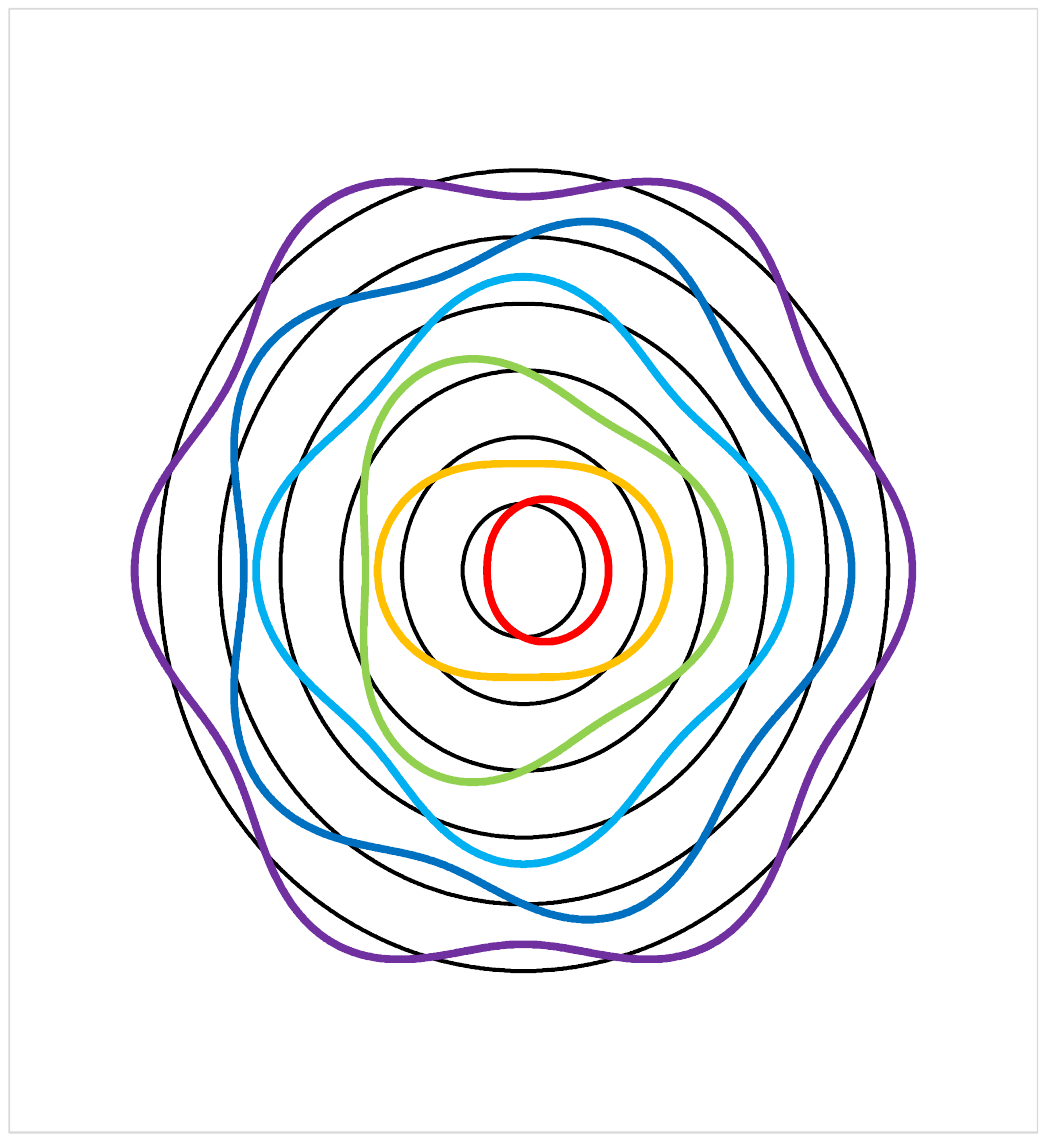}
    \caption{ ({\it rose model})shows
the pictorial presentation of the resonance wave state inside very small area.
The color of each wave corresponds to the number of waves
$n$ from 1 (red), 2 (orange), 3 (green), 4 (light blue), 5 (blue), and 6 (violet) respectively.}
    \label{fig:04a}
\end{minipage}
\begin{minipage}[t]{0.49\columnwidth}
    \centering
    \includegraphics[width=1.1\columnwidth]{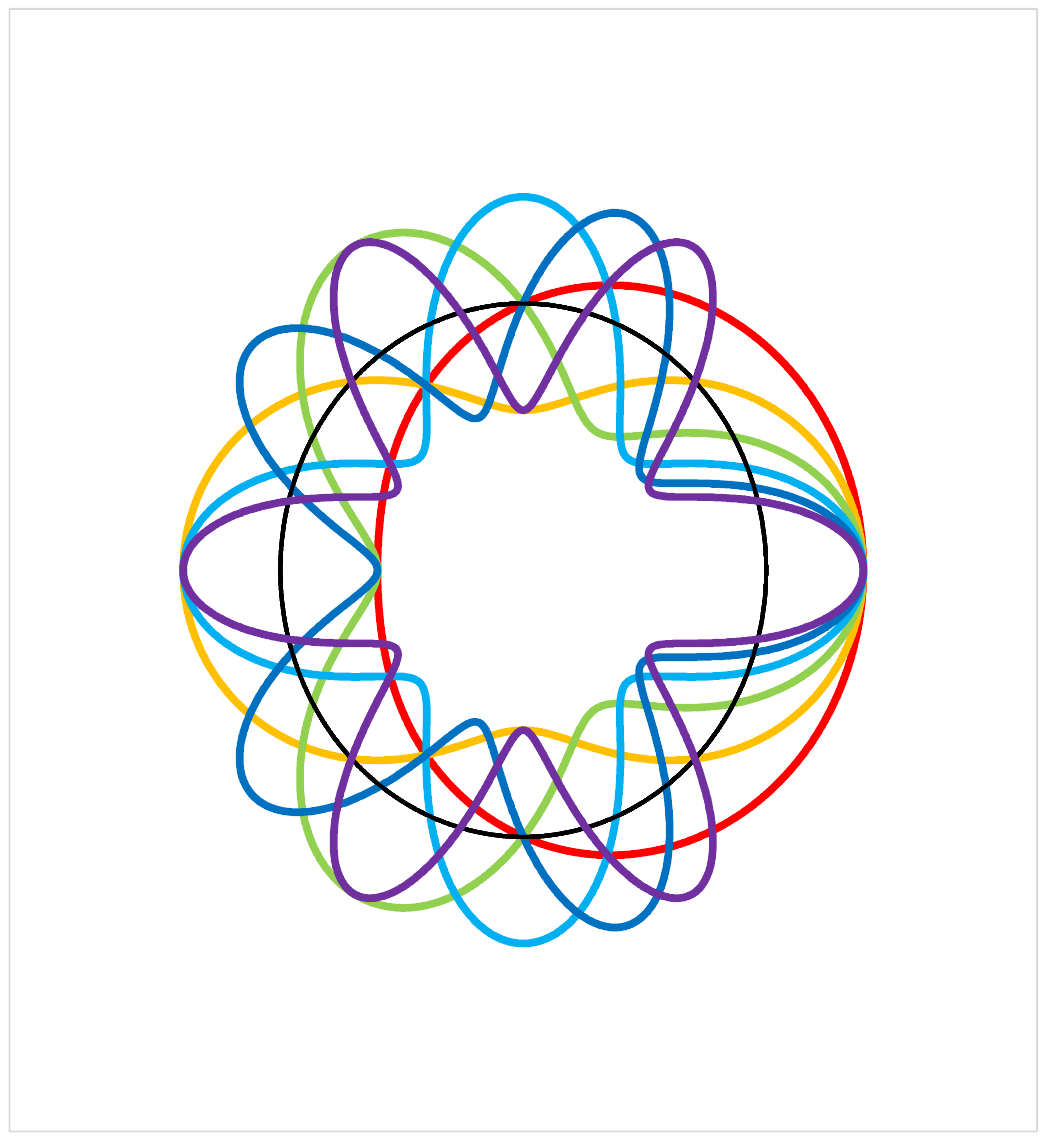}
    \caption{(
{\it sun flower model};
the number of waves is increasing on the same circle.
Our model is consistent with the rose model.
The unit length is estimated as to be
$\ell_{0}\approx 25\,{\rm fm}$ or 810 MeV.
In the case of the sun flower model,
the resonances occur in the same narrow region,
and the $v$ is not constant but is multiplied by the integer number.
Therefore, the energy level of the resonances may not show
the logarithmic spacing.
According to the solution of the Schr\"{o}dinger equation in the two dimensional plane, the energy level is expected as to be $n^{2}$ spacing.}
\label{fig:04b}
\end{minipage}
\end{figure}

\begin{figure}[t]
\centering
\includegraphics[keepaspectratio, scale=0.6]{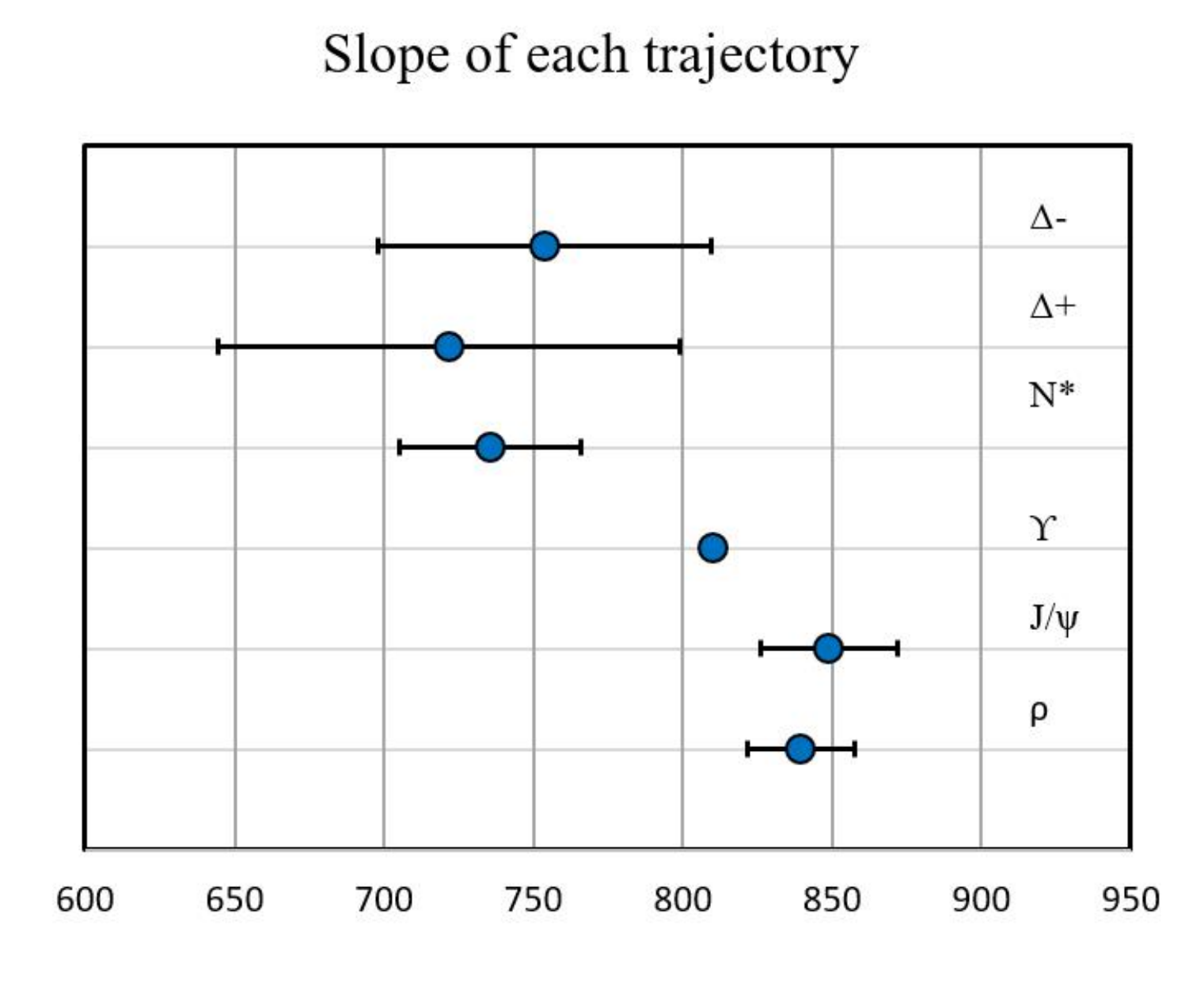}
\caption{The slope of each resonance trajectory is compared. The error bars of the baryon resonances are large, however between the slopes of $N^{*}$ and $\Upsilon$ resonances, the 3 $\sigma$ difference is recognized.}
\label{fig:05}
\end{figure}

\begin{figure}[t]
\centering
\includegraphics[keepaspectratio, scale=0.6]{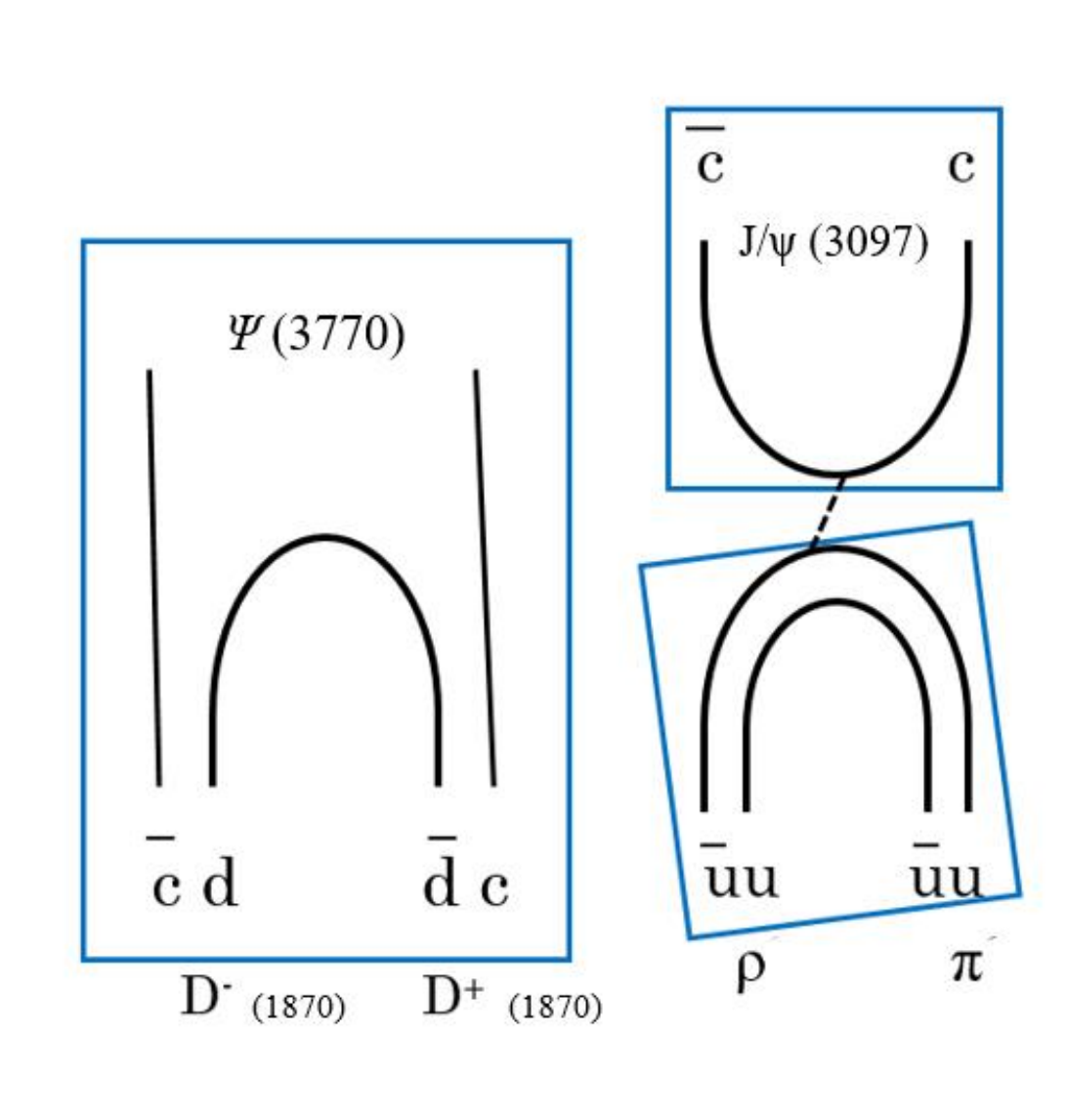}
\caption{The decay of charmonium J/$\psi$ into $\rho$ and $\pi$ mesons (right) and $\psi$ (3770) (left) are shown on the plane with the quark line. The decay rate of the right side is limited due to the OZI rule.  In our model, the OZI rule may be interpreted as the connection problem between the two planes.}
\label{fig:06}
\end{figure}

\begin{figure}[t]
\centering
\includegraphics[keepaspectratio, scale=0.4]{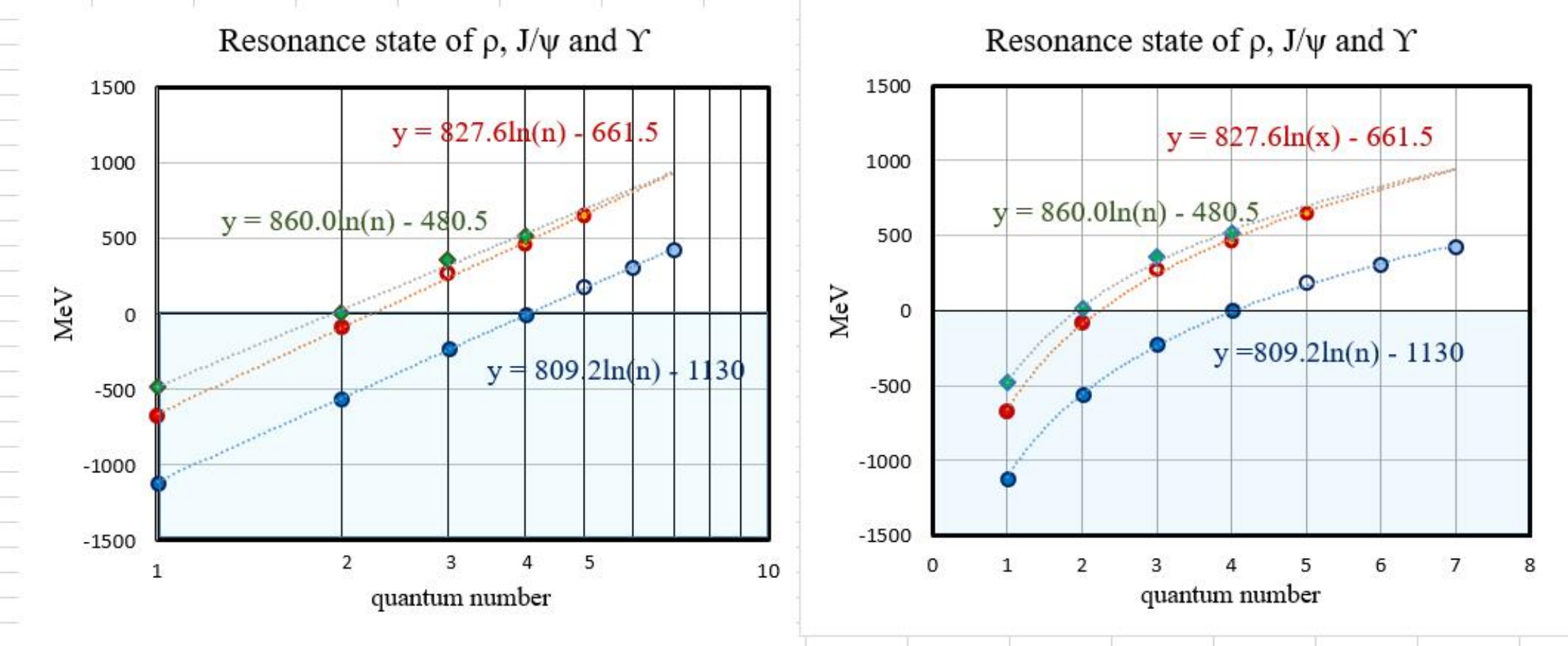}
\caption{The resonance states of $\rho$ meson series (green), charmonium (red) and botomonium (blue) are plotted on the semi-logarithmic scale diagram (left).
The open circles correspond to the states that peak energy is not well determined with narrow width.
The right side picture shows the same plots but on the linear scale for the distance $r$.  The blue area represents the deeply bounded states below th flavor threshold.}
\label{fig:07}
\end{figure}

\begin{figure}[t]
\centering
\includegraphics[keepaspectratio, scale=0.4]{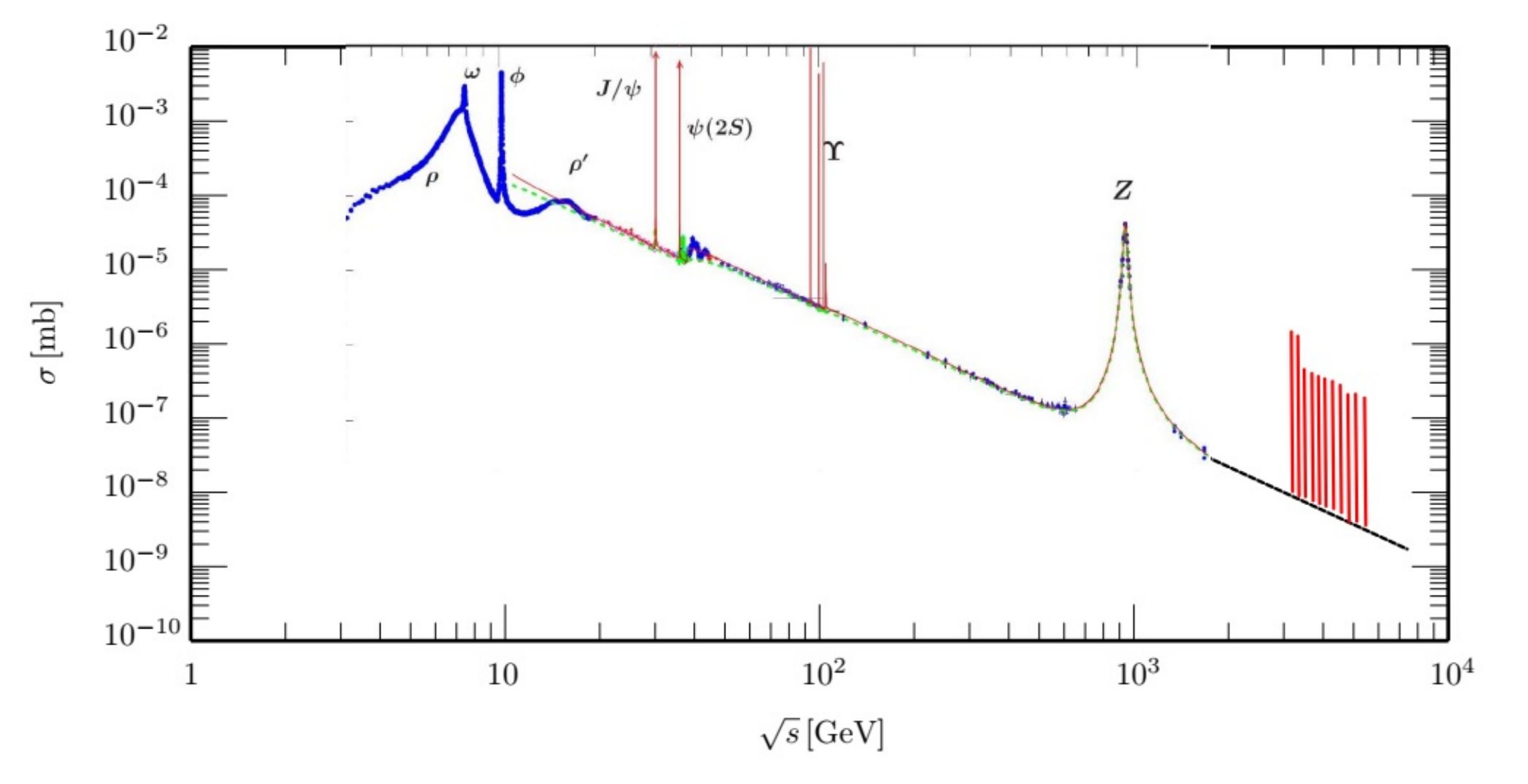}
\caption{The total cross section of $e^{+}e^{-}\rightarrow$hadrons as a function of collision energy $\sqrt{s}$ [GeV].
Possible production of toponium resonance states are also plotted over 347 GeV to 525 GeV by the red bars.}
\label{fig:08}
\end{figure}

\section*{Appendix I}

In the paper by Atabek, Deutsch, and lavaud, the Schr\"odinger equation in the two-dimensional
space is discussed under the logarithmic potential;
$\Delta$$\psi$ + [$\lambda$ - V(r)] $\psi$= 0 and V(r) = $q^{2}$ ln(r).  
By $\psi$(r,$\theta$) = f(r) and g($\theta$),
separating the coordinates into radial and polar components, they introduce the following two equations
g"($\theta$)/g($\theta$)= - $m^{2}$ and g(($\theta$))= sin(m$\theta$) or cos(m$\theta$)
    and,
for the radial direction, 
f"(r) + f'(r)/r + ($\lambda$ - V(r) - $m^{2}$/r) f(r) = 0.
For m = 0, the equation is expressed as
f"(r) + f'(r)/r + ($\lambda$ - V(r)) f(r) = 0.

Rewriting the above equation with the new coordinate f(r)=1/$\surd$r $\cdot$$\chi$(r), one obtains the new equation
 $\chi$"(r) + ($\chi$ - V(r) - ($m^{2}$-1/4)/$r^{2}$)$\chi$(r) = 0.

Compared to the $\ell$($\ell$+1)/$r^{2}$ terms that appears in the three-dimensional Shcr$\ddot{o}$dinger equation, the corresponding terms is ($m^{2}$-1/4). The term ($m^{2}$-1/4)/$r^{2}$ may correspond to the centrifugal force in the two-dimensional space.  As the particle approaches to the center (r $\sim$ 0), a strong repulsive force appears, so that even at m=0, the 1/4 term still remains.  This may correspond to the 1/2 term of the harmonic oscillator that arises from the uncertainty relation inherent to quantum mechanics. 

In the family of $\rho$-mesons, the correlation factor of {$\it R$} deviates slightly from the logarithmic formula in comparison with the J/$\psi$ and Upsilon series; the correlation coefficient {$\it R$} deviates from {$\it R$} $\sim$ 1.0 in the logarithmic formula to the $\rho$ meson family as 0.9987. This may arise from the above effect of 1/4$r^{2}$ term to the $\rho$ meson family due to the centrifugal force.  Indeed, when the resonant states of the $\rho$ meson family are plotted on a fit starting from $\it ln$(2), the correlation coefficient improves from 0.9987 to 0.99988.

 The wave function is also shown in Atabek's paper.   For the wave function of m=0, the probability amplitude shows that the particles are concentrated in the one thin ring of the  doughnut-shaped, while for m=1 (m $\neq$ 0), the probability amplitude shows two rings and three rings for m=2, as they are expected.
\end{document}